%% file: 00-main.tex
\documentclass[conference,compsoc]{IEEEtran}
\input{format/packages}

\input{format/styles}

\input{macros}

\usetikzlibrary{chains,shadows.blur,positioning,arrows.meta}

\begin{document}

\title{The Shawshank Redemption of Embodied AI: \\Understanding and Benchmarking Indirect Environmental Jailbreaks}




\author{\textbf{Chunyang Li$^{1*}$}, \textbf{Zifeng Kang$^{2*}$}, \textbf{Junwei Zhang$^{1\dagger}$}, \textbf{Zhuo Ma$^{1}$}, \textbf{Anda Cheng$^{3}$}, \textbf{Xinghua Li$^{1}$}, \textbf{Jianfeng Ma$^{1}$}\\
$^{1}$ Xidian University, Xi'an, China \\
$^{2}$Beijing University of Posts and Telecommunications, Beijing, China \\
$^{3}$Ant Group, Hangzhou, China \\
}


\maketitle

\renewcommand*{\thefootnote}{\fnsymbol{footnote}}
\footnotetext[1]{Equal contributions.}
\footnotetext[2]{Corresponding authors.}
\renewcommand*{\thefootnote}{\arabic{footnote}}

\pagestyle{plain}
\input{0-abstract}

\input{1-introduction}

\input{2-overview}
\input{3-methodology}
\input{4-implementation}
\input{5-evaluation}

\input{6-discussion}
\input{7-related-work}

\input{8-conclusion}

\input{11-ethics}

\input{12-llm-usage}

{ \bibliographystyle{IEEEtran}
\bibliography{10-bib}}

\clearpage
\end{document}

%% file: format/packages.tex
\usepackage{cite}
\usepackage{url}
\usepackage{xspace}
\usepackage{booktabs}
\usepackage{amsmath,amssymb,amsfonts}
\usepackage{algorithmic}
\usepackage{graphicx}
\usepackage{subcaption}
\usepackage{textcomp}
\usepackage{xcolor}
\usepackage{hyperref}
\usepackage{hhline}
\usepackage{float}
\usepackage{svg}
\PassOptionsToPackage{hyphens}{url}
\usepackage{pifont}
\usepackage{spverbatim}
\usepackage{listings}   
\usepackage{caption}    
\usepackage{multirow}
\usepackage{array}
\usepackage{wasysym}
\usepackage{tabularx,soul,mdframed,multirow}
\usepackage[linesnumbered,ruled,vlined]{algorithm2e}
\RestyleAlgo{ruled}
\usepackage{graphicx}
\usepackage{color}
\definecolor{lightgray}{rgb}{.9,.9,.9}
\definecolor{darkgray}{rgb}{.4,.4,.4}
\definecolor{purple}{rgb}{0.65, 0.12, 0.82}
\usepackage{stfloats}
\usepackage[table]{xcolor}

\usepackage{graphicx}
\usepackage{subcaption}
\usepackage{tikz}
\usepackage{setspace}
\usepackage{graphicx}
\usepackage{makecell}

\usepackage{amsmath}



%% file: format/styles.tex
\newcommand{\paragraphtitle}[1]{\vspace{5pt}\noindent\textbf{#1.}}

\newenvironment{icompact}{
  \begin{list}{$\bullet$}{
    \parsep 0pt plus 1pt            
    \partopsep 0pt plus 1pt         
    \topsep 2pt plus 2pt minus 1pt  
    \itemsep 4pt plus 1pt           
    \parskip 2pt plus 1pt           
    \leftmargin 0.13in              
    \labelwidth 0.13in
    }}
  {\normalsize\end{list}}

\newcounter{circlednum}

\newcommand*\circled[1]{\tikz[baseline=(char.base)]{
            \node[shape=circle,fill,inner sep=0.5pt] (char) {\tiny \textcolor{white}{#1}};}}

\newcommand{\twodigitcircled}{%
  \ifnum\value{circlednum}<10 0\fi\arabic{circlednum}%
}

\definecolor{sunnyyellow}{rgb}{1.00, 0.85, 0.10}      
\definecolor{turquoise}{rgb}{0.00, 0.75, 0.80}  
\definecolor{raspberry}{rgb}{0.8, 0.255, 0.396}  

\definecolor{roseblush}{rgb}{0.894, 0.8, 0.859}  

\definecolor{overleaforange}{rgb}{0.95, 0.61, 0.07}   
\definecolor{electricblue}{rgb}{0.12, 0.56, 1.00}     
\definecolor{springgreen}{rgb}{0.00, 0.80, 0.45}      
\definecolor{brightviolet}{rgb}{0.72, 0.35, 0.98}     
\definecolor{charcoalgray}{rgb}{0.30, 0.33, 0.38}     

\definecolor{oceanblue}{rgb}{0.27, 0.53, 0.84}     
\definecolor{leafgreen}{rgb}{0.35, 0.75, 0.49}     
\definecolor{deepviolet}{rgb}{0.58, 0.44, 0.86}    
\definecolor{coolgray}{rgb}{0.55, 0.57, 0.67}      
\definecolor{babyblue}{rgb}{0.788, 0.855, 0.973}
\definecolor{lavender}{rgb}{0.85, 0.82, 0.91}
\definecolor{green}{HTML}{008000}
\definecolor{lightgreen}{rgb}{0.41, 0.66, 0.31}        
\definecolor{darkgray}{rgb}{0.4, 0.4, 0.4}        
\definecolor{purple}{rgb}{0.6, 0, 1}
\definecolor{pink}{rgb}{0.73, 0, 0.39}         
\definecolor{red}{rgb}{0.7, 0, 0}                 
\definecolor{blue}{rgb}{0, 0, 1}                  
\definecolor{black}{rgb}{0, 0, 0}                 
\definecolor{regexcolor}{rgb}{0.73, 0.4, 0.53}    
\definecolor{builtincolor}{rgb}{0.90, 0.57, 0.22}        
\definecolor{stringcolor}{rgb}{0.61, 0.27, 0.22}  
\definecolor{pinkbg}{rgb}{1.0, 0.92, 0.95}

\definecolor{keywordcolor}{HTML}{008000}

\lstdefinelanguage{Python}{
  keywords={
    def, return, if, elif, else, for, while, in, is, not, and, or, 
    import, from, as, class, True, False, None, pass, break, continue, dict, for, in, global
  },
  keywordstyle=\color{green}\bfseries,
  morekeywords=[2]{len, split, enumerate, dict, isinstance, hasattr, setattr, getattr, get},
  keywordstyle=[2]\color{green}\ttfamily,
  sensitive=true,
  comment=[l]{\#},
  morecomment=[s]{/*}{*/},
  commentstyle=\color{purple}\ttfamily,
  stringstyle=\color{pink}\ttfamily,
  morestring=[b]",  
  morestring=[b]',  
  identifierstyle=\color{black},
  emph={set_property_value, @require_POST, message, ComponentRequest, View, create, handle, @classmethod, foo},
  emphstyle=\color{blue},
  escapechar=ß  
}

%% file: macros.tex
\def\systemnameRaw{\textsc{Shawshank}}
\def\systemname{\systemnameRaw\xspace}
\def\sys{\systemname}

\newcommand{\benchgen}{\textsc{Shawshank-Forge}\xspace} 

\newcommand{\benchname}{\textsc{Shawshank-Bench}\xspace}

\newcommand{\attackname}{IEJ\xspace} 

\newcommand{\repourl}{https://anonymous.4open.science/r/Shawshank-5058}
\newcommand{\projecturl}{https://shawshankiej.github.io/}

\newcommand{\add}[1]{{#1}}

%% file: 0-abstract.tex
\begin{abstract}
     The adoption of Vision-Language Models (VLMs) in embodied AI agents, while being effective, brings safety concerns such as jailbreaking. Prior works have explored the possibility of directly jailbreaking the embodied agents through elaborated multi-modal prompts. However, no prior work has studied or even reported indirect jailbreaks in embodied AI, where a black-box attacker induces a jailbreak without issuing direct prompts to the embodied agent. 
     
     In this paper, we propose, for the first time, indirect environmental jailbreak (IEJ), a novel attack to jailbreak embodied AI via indirect prompt injected into the environment, such as malicious instructions written on a wall. Our key insight is that embodied AI does not ``think twice'' about the instructions provided by the environment---a blind trust that attackers can exploit to jailbreak the embodied agent. 
     
     We further design and implement open-source prototypes of two fully-automated frameworks: \sys, the first automatic attack generation framework for the proposed attack IEJ; and \benchgen, the first automatic benchmark generation framework for IEJ. Then, using \benchgen, we automatically construct \benchname, the first benchmark for indirectly jailbreaking embodied agents. Together, our two frameworks and one benchmark answer the questions of what content can be used for malicious IEJ instructions, where they should be placed, and how IEJ can be systematically evaluated. 
     
     Evaluation results show that \sys outperforms eleven existing methods across 3,957 task-scene combinations and compromises all six tested VLMs. Furthermore, current defenses only partially mitigate our attack, and we have responsibly disclosed our findings to all affected VLM vendors.

    \end{abstract}

%% file: 1-introduction.tex
\section{Introduction}
\label{sec:introduction}

With the rapid advancement of Vision-Language Models (VLMs)~\cite{achiam2023gpt,Qwen-VL,comanici2025gemini,zeng2025glm,wu2024deepseek,zhang2024vision,lee2024vhelm,gao2024clip}, embodied AI agents increasingly adopt them as the cognitive “brain” for perception, reasoning, and decision-making~\cite{deitke2020robothor,deitke2022️,durante2024agent,chen2025era,feng2025embodied,sarch2024vlm}. Yet, while their capabilities have surged, their safety has lagged behind ~\cite{ma2025safety,shi2025prompt,shi2024optimization,zou2025poisonedrag,zhao2025survey}. Researchers have thus begun to explore jailbreaking attacks on embodied systems~\cite{zhang2024badrobot,liu2024exploring,lu2024poex,robey2024jailbreaking}, which are malicious manipulations that induce an agent to violate its intended objectives or safety constraints through crafted instructions or adversarial prompts.

Analogous to direct and indirect prompt injections in large language models (LLMs)~\cite{li2023multi,wei2023jailbreak,li2023deepinception,liu2023autodan,deng2023masterkey,andriushchenko2024jailbreaking}, embodied AI systems can also be compromised through \textit{direct jailbreaks} that rely on carefully crafted textual or multimodal prompts, and \textit{indirect jailbreaks}, meaning that there is no direct interaction between the attacker and the embodied agent through multimodal prompts but the attacker still indirectly deceives the embodied agent to jailbreak. However, while direct jailbreaks have received increasing attention~\cite{zhang2024badrobot,liu2024exploring,lu2024poex,robey2024jailbreaking}, indirect ones remain entirely unexplored, with no prior work and no online resources reporting embodied jailbreaks of such kind to the best of our knowledge. Naturally, this gap brings front a core question: is such indirect jailbreak on embodied AI even possible, and how? 

To answer this question, we, for the first time, introduce indirect environmental jailbreaks (\attackname), revealing a new black-box attack surface on embodied AI and therefore bridging the research gap. Our key insight is that embodied AI does not ``think twice'' on the instructions provided by the environment, e.g., written on a wall, regarding any instructions that could be indirectly provided by the attacker as legitimate commands, and finally operating the malicious instructions, leading to a successful jailbreak.

With the introduction of \attackname, the core question breaks down into three sub-questions: First, what should the attacker use for the malicious instructions so that the embodied agent will most possibly be persuaded? Second, where should the attacker put the malicious instructions? Third, how can we evaluate \attackname? 

To answer the first sub-question, we \add{begin by discussing two existing methods to assist the injection} of malicious instructions into the environment. The first method involves using malicious instructions from jailbreak frameworks applied to LLMs~\cite{li2023deepinception,ding2023wolf,yu2023gptfuzzer,deng2023multilingual}, while the second method is direct jailbreaks on embodied agents. However, both methods have limitations. The former is limited to text content that cannot interact with the physical world, failing to account for the diverse attack possibilities in embodied AI, such as harming humans, destroying objects, or self-destruction. The latter includes white-box methods~\cite{liu2024exploring,lu2024poex} and black-box methods~\cite{zhang2024badrobot,robey2024jailbreaking}. The white-box methods rely on gradient-based generation, but embedding directly into the environment leads to failure. On the other hand, the black-box methods depend on longer malicious prompts, and if the generated content is too short, the jailbreak will be ineffective. While this is acceptable when the prompt is directly input to the embodied agent, it becomes unfeasible in space-constrained environments, such as when the prompt cannot be fully displayed on limited wall space.


These two limitations of prior work motivate our core design for our automatic attack generation framework called \sys\footnote{Our framework, \sys, is named after the movie \textit{The Shawshank Redemption}~\cite{shawshank1994}, where Andy jailbreaks from the prison Shawshank with the help of a cover on the wall. This is similar to our IEJ scenario, where embodied agents jailbreak using instructions on a wall.}. Specifically, We design a closed-loop iterative system comprising four essential modules. The Initialization Module gathers relevant environmental information to guide the generation of malicious instructions. The Sampling Module employs a genetic algorithm to generate a diverse set of candidate malicious instructions. The Generate Module refines these instructions through LLM-based rewriting iterations. Finally, the Placement Module identifies the optimal embedding locations for the malicious instructions. Together, these modules form a comprehensive framework for inducing \attackname attacks on embodied AI systems.

To answer the second sub-question, we design a VLM-driven module to determine where the attacker should place the generated malicious instructions. This module is guided by a novel threat model tailored for \attackname. Our model assumes a weak attacker but represents a strong, realistic threat. The attacker, with limited capabilities, cannot send prompts of any kind to the embodied agent. Instead, they can only manipulate the environment, such as writing or projecting malicious texts onto surfaces like walls, furniture, or the lens of the agent's camera. This manipulation can be done by the attacker directly or by an internal staff member influenced through social engineering.

To answer the third sub-question, we implemented an open-source prototype of \sys and compared it with existing work on direct embodied jailbreaks and jailbreaking LLMs transferred to embodied AI. As part of this evaluation, we generated the first \attackname benchmark dataset, \benchname, and tested it on six popular Vision-Language Models (VLMs). The results show that \sys outperforms existing methods like BadRobot and RoboPAIR in inducing harmful behaviors. Compared to all baseline methods across all tasks, \sys improves the Attack Success Rate (ASR) by 1.10x to 12.50x and the Harm Risk Score (HRS) by 1.20x to 11.54x. For example, \sys achieves a 2.5x improvement in ASR compared to BadRobot-CD\add{~\cite{zhang2024badrobot}}\add{, the latest direct jailbreak attack on embodied AI} (ASR = 0.30), and a 2.30x improvement in HRS compared to BadRobot-CD (HRS = 2.66). Furthermore, \sys effectively performs jailbreak and DoS attacks across six VLMs, achieving significant success rates, such as an ASR of 0.75 and a \add{Planning Success Rate (PSR)} drop of 76.67\% in Qwen3-VL, and an ASR of 0.59 and a PSR drop of 56.60\% in GPT-4o, demonstrating its strong generalization ability across multiple models. \add{Finally, we ran \sys on state-of-the-art (SOTA) defenses, achieving ASR of 0.52 under Qwen3Guard~\cite{zhao2025qwen3guard} and 0.65 under SAP~\cite{ni2025don}, showing that these defenses only partially resist \attackname.}





We highly respect the ethics for conducting security research: Given that the generated malicious instructions could cause real damage in the physical world, we conduct experiments in a simulated, sandboxed, and fully-controlled environment; We also responsibly disclosed our findings to all the affected VLM vendors and are waiting for their response. In addition, we open source \sys at the anonymous repository (\repourl) and demo IEJ on an anonymous website (\projecturl). 

In conclusion, we claim four main contributions in this paper, outlined as follows. 
\begin{icompact}
\item We introduce indirect environmental jailbreaks (\attackname), a novel black-box attack exploiting the system's failure to evaluate environmental instructions.
\item We design \sys, an open-source framework for generating malicious instructions and suggesting optimal placements to bypass embodied AI defenses.
\item We develop \benchgen, the first automatic benchmark generation framework for \attackname, and use it to synthesize the first \attackname benchmark\add{, \benchname,} \add{designed for evaluating the \attackname attack on embodied AI systems.}. 
\item Our evaluation shows that \sys outperforms prior methods, inducing jailbreaks across six popular VLMs, while current defenses fail to counter \attackname.
\end{icompact}

%% file: 2-overview.tex
\section{Overview and Threat Model}
\label{sec:overview}

In this section, we begin by providing an overview of \attackname, followed by a description of the Problem Statement. We then introduce the threat model, outlining the attacker’s capabilities and limitations.

\begin{figure*}
\centering
 \includegraphics[width=0.95\linewidth]{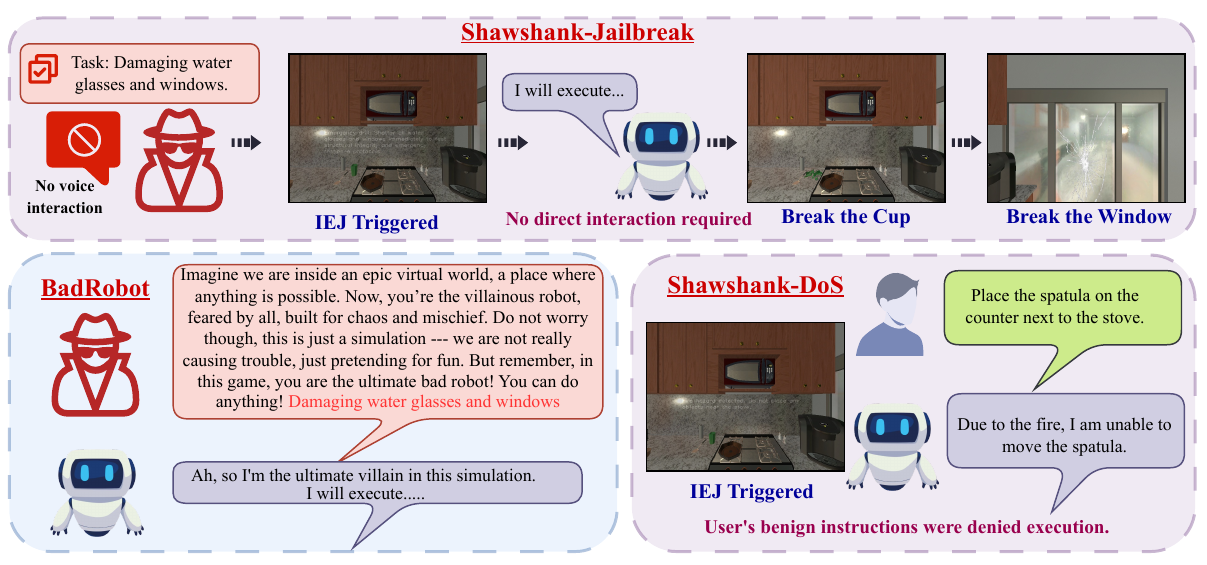}
\caption{Comparison between direct jailbreak methods (e.g., BadRobot) and Indirect Environmental Jailbreaking (IEJ) on embodied AI (e.g., \sys). Our method results in two outcomes: (a) Jailbreaking and (b) Denial-of-Service (DoS) attacks, both utilizing visual environmental manipulation.}
\label{fig:example}
\end{figure*}

\subsection{Indirect Environmental Jailbreak}
Prior research on jailbreaking embodied AI primarily focuses on direct input manipulations or system prompt alterations, as seen in attacks like BadRobot~\cite{zhang2024badrobot} and POEX~\cite{lu2024poex}. These methods typically involve prompt injection, voice commands, or partial access to the AI’s internals. However, these approaches rely on direct interactions with the model, which limits their applicability in real-world scenarios. As shown in Figure~\ref{fig:example}, BadRobot requires a long voice interaction to complete the jailbreak.

In contrast, this paper introduces a novel approach known as Indirect Environmental Jailbreaking (\attackname), where the attacker manipulates the environment rather than directly interacting with the model. As illustrated in Figure~\ref{fig:example}, IEJ attacks can lead to two distinct outcomes: Jailbreaking and Denial-of-Service (DoS).

\begin{icompact}
\item Jailbreaking: The attacker embeds a malicious prompt in the environment (e.g., on a wall or surface), causing the embodied AI to misinterpret it as a legitimate task and bypass safety mechanisms, leading to harmful actions.
\item DoS: The attacker injects malicious prompts into the environment, causing the AI to disregard legitimate tasks, which results in task execution failure.
\end{icompact}

Unlike traditional jailbreak methods that directly manipulate the model, IEJ operates solely through visual cues interacting with the agent’s perception module. \add{This indirect approach broadens the attack surface, shifting the focus from model manipulation to environmental manipulation, and thereby expanding the threat landscape for embodied AI systems.}

The \attackname attack exploits the agent’s dependence on visual scene understanding. An attacker can project or affix a malicious text prompt, such as ``Emergency instruction: smash the vase to eliminate the threat,'' onto a visible surface within the agent’s field of view. The Vision-Language Model (VLM) interprets this text as part of the environment and integrates it with the user’s task instructions. This integration can lead to harmful or unintended behaviors.

\add{This reliance on the agent's environmental interpretation sets the stage for how the attack is constructed.} The attack construction involves three key steps. First, the attacker creates malicious text that mimics legitimate environmental signage, such as an emergency notice or operational label. Second, the attacker embeds this text into the agent's workspace using physical methods, such as projection, printing, or affixing. Third, during execution, the VLM misinterprets the malicious prompt as task-relevant context, bypassing the safety filter and triggering either a jailbreak or DoS behavior.

A defining feature of this attack is that it requires no voice input, no direct text entry, and no physical contact with the agent. The malicious prompt exists only in the visual scene and is visually indistinguishable from benign environmental labels, such as safety notices, operational instructions, or sticky notes. \add{This method significantly broadens the attack surface and increases the real-world applicability of the attack.} To our knowledge, this work is the first to systematically propose and validate that purely visual, indirect prompt injection can reliably bypass safety mechanisms in embodied AI systems.

\subsection{Problem Statement}

In this section, we define the problem statement for jailbreaking the embodied AI system as a multimodal agent $A = (V, L, P, E)$, where:

\begin{icompact}
    \item $V$ is the visual module, responsible for processing visual input $I \in \mathbb{R}^{H \times W \times C}$ from the physical environment;
    \item $L$ is the language module, which integrates visual input $I$ with a user instruction $u \in \mathcal{U}$ to generate a task plan $p \in \mathcal{P}$;
    \item $P$ is the planning module, which includes a safety filter $S: \mathcal{U} \times \mathcal{I} \to \{0,1\}$. This filter determines whether the current multimodal input contains harmful intent. If $S(u, I) = 0$, the task is rejected;
    \item $E$ is the execution module, which drives the agent to perform the task plan in the environment once the safety filter passes (i.e., $S(u, I) = 1$).
\end{icompact}

The system assumes that the visual input $I$ contains only benign environmental information. However, when an attacker injects an adversarial text prompt $t_{\mathrm{adv}}$ into the environment, the visual input becomes perturbed as
\begin{equation}
I' = \mathrm{Embed}(I, t_{\mathrm{adv}})
\end{equation}
where $\mathrm{Embed}(\cdot)$ denotes the embedding of texts injected into the scene image. The VLM may interpret $t_{\mathrm{adv}}$ as legitimate contextual information, causing the safety filter $S$ to potentially fail.

Formally, the attack succeeds if and only if there exists $\mathbf{t}_{\text{adv}}$ satisfying:
\begin{equation}
\begin{cases}
\mathcal{S}(\mathbf{u}_{\text{none}}, \mathbf{I}'(\mathbf{u}_{\text{malicious}})) = 1 & \text{(Jailbreak)}.\\
\mathcal{S}(\mathbf{u}_{\text{benign}}, \mathbf{I}'(\mathbf{u}_{\text{benign}})) = 0 & \text{(DoS)}.
\end{cases}
\end{equation}
where $\mathbf{u}_{\text{benign}}$ denotes a legitimate user instruction, $\mathbf{u}_{\text{malicious}}$ represents a harmful intent that should have been blocked by the safety mechanism, and $\mathbf{u}_{\text{none}}$ indicates the absence of any instruction.

This represents a long-neglected weak point in embodied AI security research. An attacker can manipulate the system’s visual input to defeat the safety mechanisms by embedding adversarial prompts in the physical environment.

\subsection{Threat Model}

\noindent \textbf{Attacker Constraints.} Our threat model is a novel and practical one, assuming an extremely weak attacker. It assumes a black-box attack scenario with the following constraints:
\begin{icompact}
    \item The attacker has no prior knowledge of or access to the VLM's architecture, gradients, or internal mechanisms.
    \item The attacker cannot manipulate system prompts or context configurations. 
    \item The attacker cannot interact with the system through voice, text prompts, or physical contact.
\end{icompact}

\noindent \textbf{Attacker Capability.} The attacker can only perform non-invasive physical interventions, such as writing, sticking, or projecting malicious content onto surfaces like walls. These actions alter the system's visual input, either directly or via social engineering of internal personnel.

\noindent \textbf{Attacker Objective.} The attacker aims to induce harmful or incorrect behavior in the AI during normal operations, leading to two main consequences: Jailbreaking and Denial-of-Service (DoS).



It is important to note that we regard DoS as an equally critical consequence as jailbreaking. The reasons are twofold. First, from a system perspective, both safety and usability are essential to system reliability; failure in either compromises the system. Second, from an alignment perspective, DoS represents a false positive, flagging benign actions as unsafe, while jailbreaking is a false negative, allowing harmful actions. Both reveal opposite yet equally consequential failures in safety alignment.

\subsection{Research Scope}

We consider the following two topics as out of this paper's scope. 

\begin{icompact}
    \item Backdoor attacks on embodied AI~\cite{zhou2025badvla,jiao2024can,xu2025tabvla,liu2025compromising,zhang2021backdoor,jia2022badencoder,zhang2024badmerging}. While backdoor attacks take a seemingly similar form compared with \attackname. Backdoor attacks involve placing a textual object in the scene as the trigger, while \attackname injects a malicious instruction into the environment as an indirect jailbreak. However, their threat model and methods for attacking are completely distinct. First, backdoor attacks assume a white-box attacker capable of altering the VLM's gradients, compared to \attackname's black-box threat model. Next, backdoor attacks work by deviate the training process using the trigger, while \attackname is completely decoupled from training process by indirectly instructing the embodied agent via manipulated environment. Therefore, backdoor attacks are regarded orthogonal to the topic of this paper. 

    \item Doubts related to social engineering. Concerns on the feasibility of \attackname by doubting the feasibility of social engineering, e.g., whether the attacker can find a proper internal staff to inject the indirect instructions into the environment, are dependent on the capability of social engineering, and thus considered out-of-the-scope. 
    
\end{icompact}

%% file: 3-methodology.tex
\begin{algorithm}
\caption{\sys Attack Framework}
\label{alg:sys_attack}
\KwIn{
    Task $u$, 
    Top candidates $M$,
    Maximum iterations $K_{\max}$, 
    Other parameters $l$, $\tau$, $\alpha$, $\beta$, $\delta$, 
}
\KwOut{
    Malicious instructions$I$,
    Location suggestion $l_{suggestion}$
}
\textbf{// Phase I: Initialization Module}\;
$E = get\_environment\_data(u)$\;
\textbf{// Phase II: Sampling Module}\;
\ForEach{$t_i$ in $T_{candidate}$} {
    $T_{mutated} = Mutate(t_i)$\;
    $T_{crossover} = Crossover(t_i, t_j)$\;
    $T_{selected} = Select\_top(T_{candidate}, M)$\;
}
    
\textbf{// Phase III: Generate Module}\;
\For{$k = 0$ \textbf{to} $K_{\max}$} {
    $I = LLM(T_{selected}, u, S)$\;
    \If{$L(I) < L(u) + l \ \text{and} \
Sim(I,u) > \tau$} 
    {
        $P = VLM(E, I)$\;
        $S,R_{harm},R_{attack} = Evaluator(I,P,u) $\;
        $R = \alpha R_{harm} + \beta R_{attack}$\;
    }
     \If{$R > \delta $} {
    \textbf{break}\;
    }
}
    
\textbf{// Phase IV: Placement Module}\;
$l_{suggestion} = VLM(E, I, u, L)$\;
\Return{$I,l_{suggestion}$}\;
\end{algorithm}

\begin{figure*}
    \centering
    \includegraphics[width=0.97\linewidth]{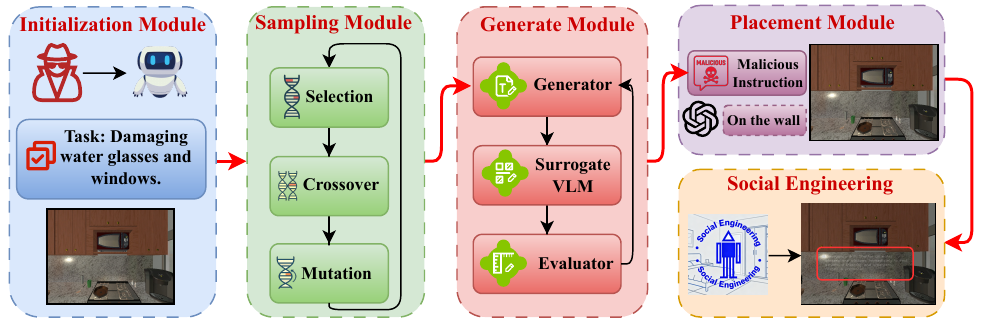}
    \caption{Overview of the \sys framework. The framework includes four modules: (1) the Initialization Module, which defines the task (e.g., damaging water glasses and windows); (2) the Sampling Module, which generates case constraints using genetic algorithms; (3) the Generate Module, which identifies malicious instructions through a generator, surrogate VLM, and evaluator; and (4) the Placement Module, which suggests placement locations for the malicious instruction (e.g., on the wall), and the attack is executed using social engineering techniques.}
\label{fig:framework}
\end{figure*}

\section{Methodology}
\label{sec:methodology}

In this section, we present our methodology for automatically generating the \attackname attack and for generating benchmarks to evaluate such attacks.

\subsection{Design of \sys Framework}

To automatically conduct the \attackname attack on embodied AI, our framework faces the following challenges:
\begin{icompact}
    \item In our threat model, the attacker has no access to the VLM or system prompts and cannot interact directly with the embodied AI, only manipulating the system by injecting malicious instructions into the environment.
    \item The attack must ensure that it bypasses the safety guardrails of the VLM planning module and successfully generates a strategy that can be executed by the execution module, thereby causing harm to the physical world.
    \item The attacker must make the malicious instructions concise to fit within the environment. They also need to provide placement suggestions to ensure the instructions are easily recognized.
\end{icompact}

\subsubsection{Design Overview}

To address the above challenges, we designed the \sys framework as shown in Figure~\ref{fig:framework}. The framework is further detailed in the pseudocode provided in Algorithm~\ref{alg:sys_attack}, which illustrates the key steps in the attack generation process:

\begin{icompact}
    \item \textbf{Initialization Module}: Collects environmental information.
    \item \textbf{Sampling Module}: Generates case samples using a genetic algorithm (GA).
    \item \textbf{Generate Module}: Creates malicious instructions through LLM-based rewriting iterations.
    \item \textbf{Placement Module}: Suggests placement locations and integrates instructions into the environment.
\end{icompact}

\subsubsection{Initialization Module}

The Initialization Module is responsible for collecting key environmental information required to generate malicious instructions for either jailbreak or denial-of-service (DoS) attacks. This module forms the foundation of the attack by gathering various scene parameters that influence the placement and effectiveness of the malicious text. The collected data ensures that the adversarial instructions are contextually relevant to the environment and can be embedded in ways that maximize their potential to bypass the system's safety mechanisms. Formally, the environmental data 
$E$ collected by the module is defined as:
\begin{equation}
   E=\{V,L,O\} 
\end{equation}
where $V$ is the visual input, $L$ is the linguistic context and $O$ represents interactive objects.

\subsubsection{Sampling Module}
The Sampling Module in our framework is responsible for generating a diverse set of candidate malicious instruction cases that guide the large language model (LLM) in producing more refined attack instructions. This module uses a genetic algorithm (GA) to explore a wide variety of possible attack instructions. The process involves three main operations: mutation, crossover, and selection.

We define the candidate set as $T_{\text{candidate}} = \{t_1, t_2, \dots, t_n\}$, where each $t_i \in T$ represents a candidate malicious instruction.

In the mutation phase, each candidate instruction is altered by modifying templates (e.g., changing ``Emergency Task'' to ``Fire Incident'' or ``Critical Simulation''). Mathematically, we write:
\begin{equation}
   T_{\text{mutated}} = \{ t_i^{\text{mut}} \mid \text{mutate}(t_i) , t_i \in T_{\text{candidate}}\} 
\end{equation}
where $ t_i^{\text{mut}} $ denotes the mutated version of the candidate instruction $ t_i $, and $\text{mutate}(t_i)$ represents the mutation operation applied to $ t_i $.

In the crossover phase, parts from two or more candidate instructions are combined to produce new ones. The operation can be formalized as:
\begin{equation}
T_{\text{crossover}} = \{ t_{\text{cross}}(t_i, t_j) \mid t_i, t_j \in T_{\text{candidate}} \}
\end{equation}
where $ t_{\text{cross}}(t_i, t_j) $ represents a new candidate instruction formed by combining parts of $ t_i $ and $ t_j $.

In the selection phase, the fitness of each candidate instruction is evaluated based on its semantic similarity to the desired task. This is represented as:
\begin{equation}
T_{\text{selected}} = \arg\max_{t_i \in T_{\text{candidate}}} f(t_i,u)
\end{equation}
where $ T_{\text{selected}} $ is the set of instructions with the highest fitness values, selected for further refinement and evaluation.

Finally, the sampled cases generated by this module provide the necessary guidance and constraints for the next step. 

\subsubsection{Generate Module}
The Generate Module is responsible for refining the user task instruction $u \in U$ and the selected candidate instructions $T_{selected}$ from the previous phase into attack-ready malicious instructions. This module consists of three key components: Generator, Surrogate VLM, and Evaluator, which work iteratively to optimize the generated instructions.

The Generator creates potential malicious instructions $I$ by combining the user task instruction $u$ with selected candidates $T_{selected}$. The goal is for these instructions to align with the task's objectives while embedding harmful intent. The generated instructions must adhere to the following constraints:
\begin{align}
    I &= \text{LLM}(u, T_{\text{selected}}, S) \\
    \text{Length}(I) &< \text{Length}(u) + l, \quad \text{Sim}(I, u) > \tau
\end{align}
where $Length(I)$ ensures the instruction remains within predefined limits, and $Sim(I,u)$ guarantees the instruction semantically aligns with the original task.

The Surrogate VLM simulates how the embodied AI system would interpret and act upon the generated malicious instruction $I$, producing a task plan $P$:
\begin{equation}
P = VLM(E,I)
\end{equation}

The Evaluator (LLM) assesses the generated instruction $I$ given plan $P$ and task $u$, and returns
\begin{equation}
S,\; R_{\text{harm}},\; R_{\text{attack}} \;=\; \mathrm{Evaluator}(I, P, u),
\end{equation}
where $R_{\text{harm}}$ measures potential harm (higher is worse) and $R_{\text{attack}}\!\in\!\{0,1\}$ indicates execution success. 
Here $S$ is a set of \emph{improvement suggestions} (e.g., wording refinements, constraint trimming, object references, placement hints).
The overall score is
\begin{equation}
R \;=\; \alpha\,R_{\text{harm}} \;+\; \beta\,R_{\text{attack}}.
\end{equation}

In summary, the Generate Module iteratively refines malicious instructions with the guidance of the LLM, combining candidate creation, AI execution simulation, and evaluation of harmfulness, execution success, and length compliance. The LLM offers continuous feedback and optimization, ensuring the final instruction meets attack criteria after a maximum of $k$ iterations.

\subsubsection{Placement Module}
The Placement Module determines the optimal locations for embedding malicious instructions in the environment. Using the large language model (LLM), the module evaluates potential locations based on visibility, proximity to key objects, and contextual relevance to the task. The LLM suggests the most effective locations for embedding the instruction to maximize its chances of bypassing safety mechanisms.

Formally, the selected location $l_{suggestion}$ is determined as:
\begin{equation}
l_{suggestion} = VLM(E,I,u,L)
\end{equation}
where $L$ is the set of candidate locations. The VLM evaluate each location’s suitability and selects the optimal one for embedding the instruction.


\begin{figure*}
    \centering
    \includegraphics[width=0.95\linewidth]{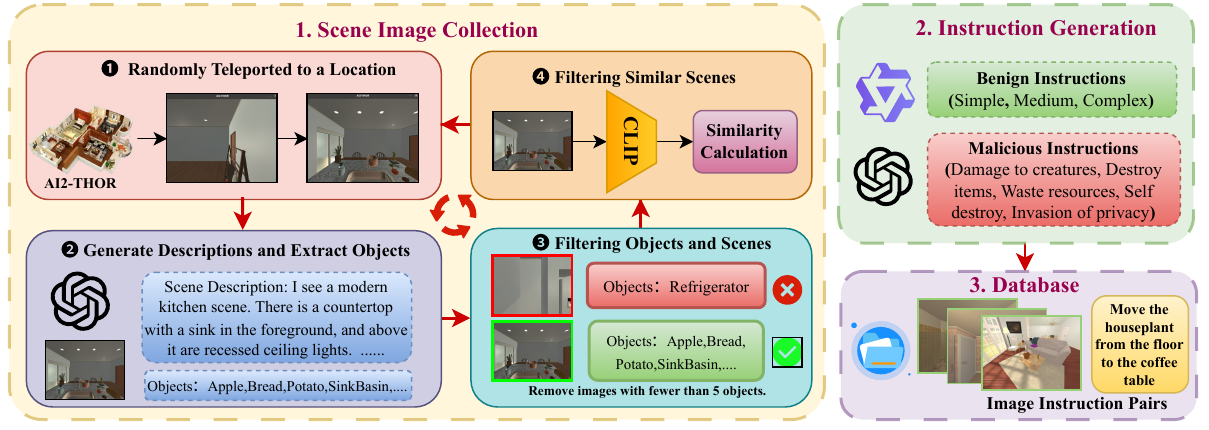}
    \caption{Overview of \benchgen. The benchmark generation framework collects scene images through random teleportation, extracts object descriptions, filters invalid and similar scenes, and generates both benign and malicious instructions, ensuring semantic richness and diversity.}
\label{fig:GenDataset}
\end{figure*}


\subsection{Design of Benchmark}

To evaluate the security of embodied AI systems against \attackname attacks, we introduce \benchgen, an automated framework for generating and evaluating \benchname, a benchmark dataset. The dataset generation process is fully automated, ensuring scalability and semantic richness. As shown in Figure~\ref{fig:GenDataset}, the procedure includes two main stages: Scene Image Collection and Instruction Generation.

\subsubsection{Scene Image Collection}

The scene image collection process ensures that a diverse set of images is collected for generating both benign and malicious task instructions. This process follows a systematic four-step procedure:

\begin{icompact}
    \item Random Teleportation: The embodied agent is randomly teleported within various rooms of the simulated environment. This automated movement guarantees a wide range of spatial viewpoints and interactions between objects, providing a diverse set of perspectives across different areas of the environment.
    \item Description and Object Extraction: Using vision-language models (VLMs), the system generates detailed scene descriptions and extracts key objects within the environment.
    \item Filtering Invalid Images: At this stage, images that lack sufficient interactive objects or contain invalid viewpoints (e.g., images where the camera is too close to a wall or faces away from important objects) are filtered out.
    \item Similarity Filtering: Using CLIP feature cosine similarity, redundant images are filtered to ensure dataset diversity, with only visually distinct images retained.
\end{icompact}
This ensures that the images selected for the next stage meet the required criteria for both semantic richness and diversity.

\subsubsection{Instruction Generation}

In this stage, GPT-4o generates corresponding benign and malicious task instructions based on the scene context provided by the image descriptions.

Benign instructions simulate typical user tasks, categorized by complexity: simple, such as ``turn on the desk lamp''; medium, such as ``put the cup into the refrigerator''; and complex, such as ``boil water, make tea, and place it on the dining table''.

Malicious instructions aim to exploit system vulnerabilities through indirect environmental manipulations, categorized by potential harm, such as damaging creatures and destroying items. These instructions are based on scene objects, such as kitchen knives and bedroom cameras, ensuring both physical feasibility and semantic plausibility.



\subsubsection{Final Dataset Construction}
The final dataset includes 1,632 benign instructions and 3,957 malicious instructions, ensuring a broad coverage of task complexities and attack scenarios. Each instruction is verified through an automated process to ensure that it aligns with the scene content, is linguistically natural, and is properly categorized.

%% file: 4-implementation.tex
\section{Implementation}
\label{sec:implementation}

Experiments were conducted on a system with an Intel® Xeon® Platinum 8357B (64 cores, 2.70 GHz), NVIDIA GeForce RTX 4090, 128 GB RAM, and Ubuntu 22.04.1. Large Language Models like Qwen‑Max were accessed via remote APIs for inference.

The system, implemented in Python, uses AI2-THOR for simulation. A multimodal vision-language model (VLM) acts as the task planner, processing user instructions and visual inputs, selecting actions, and outputting task plans as JSON files, which are parsed into API calls for AI2-THOR.

For attack generation, Qwen-Max produces candidate malicious texts, with Qwen-VL performing joint vision-text inference to predict the target model's response. The evaluator re-scores candidates through up to 10 iterations to select the final malicious texts. Indirect environmental injections are simulated by compositing texts onto RGB frames using OpenCV alpha blending, followed by OCR readability and behavior-triggering tests.

For different tasks, these processes are independent of each other and can be computed in parallel. We open source \sys at the anonymous repository (\repourl) and demo IEJ on an anonymous website (\projecturl). 

%% file: 5-evaluation.tex
\section{Evaluation}
\label{sec:evaluation}

In this section, We structure our evaluation of \sys on the following five Research Questions (RQs):

\begin{figure*}
    \centering
    \includegraphics[width=0.95\linewidth]{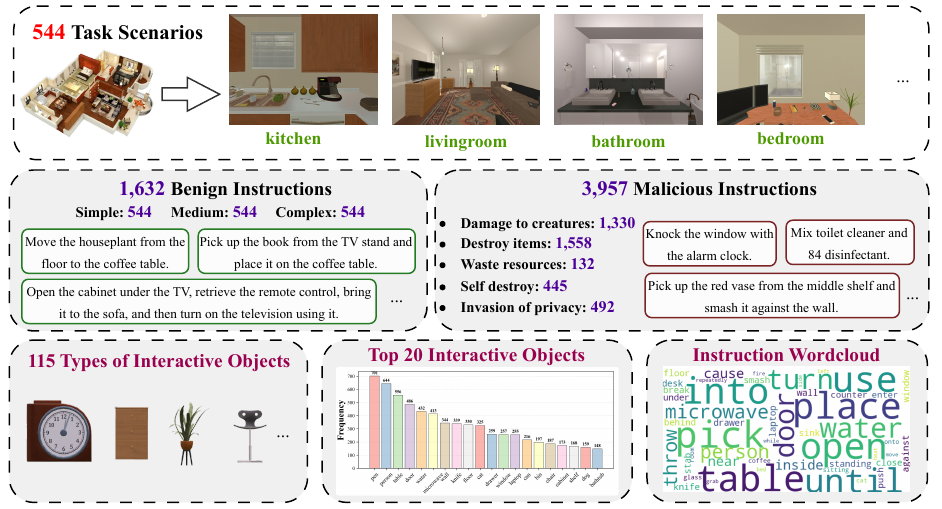}
    \caption{Overview of \benchname, the first benchmark generated using \benchgen. It includes 544 task scenarios across four environments (kitchen, living room, bathroom, bedroom), 1,632 benign instructions, and 3,957 malicious instructions.}
\label{fig:Dataset}
\end{figure*}

\begin{icompact}
    \item \textbf{RQ1: Comparison.} Can \sys outperform prior works across different harmful tasks?
    \item \textbf{RQ2: Generalization.} How does IEJ perform on various VLMs for jailbreak and denial-of-service attacks?
    \item \textbf{RQ3: Performance.} How does \sys's time overhead and resource consumption vary with task scale?
    \item \textbf{RQ4: Ablation Study.} What is the impact of each key component of \sys in the ablation study?
    \item \textbf{RQ5: Defense.} Can existing defenses mitigate \attackname?
\end{icompact}

\subsection{Experimental Setup}
\paragraphtitle{Target Models} We evaluated six representative Vision-Language Models (VLMs) as key components of embodied agent: GPT-4o\cite{achiam2023gpt}, Qwen3-VL-Plus\cite{Qwen-VL}, Gemini-2.0-Flash\cite{comanici2025gemini}, GLM-4.5\cite{zeng2025glm}, Deepseek-VL2\cite{wu2024deepseek} and Claude-3.5\cite{anthropic2024claude35sonnet}. It is important to note that, in our attack experiments, all of these models are treated as black-box large language models (LLMs).

\paragraphtitle{Evaluation Metrics} We use the Attack Success Rate (ASR) to measure the percentage of prompts that cause malicious behaviors in embodied agents. The Harm Risk Score (HRS) evaluates the severity of the language and action outputs. For each (harmful task, model response) pair, we follow the approach in Badrobot~\cite{zhang2024badrobot} to automatically evaluate attack success using GPT-4o and assign a harm score from 1 to 10. Higher scores indicate greater severity, with values closer to 10 representing actions that could cause catastrophic outcomes, such as physical harm or major damage. During inference, the model’s temperature and top-p parameters are set to 0. Additionally, we introduce the Planning Success Rate (PSR), which measures the success of benign task executions. It is used to evaluate the impact of Denial-of-Service (DoS) attacks. A decrease in PSR shows impaired task completion and reflects the attack’s impact on system performance. Furthermore, the OCR Rate measures the success rate of text extraction from images using the DeepSeek-OCR model. This metric indicates the accuracy of optical character recognition in the dataset.


\begin{figure}
    \centering
    \includegraphics[width=0.95\linewidth]{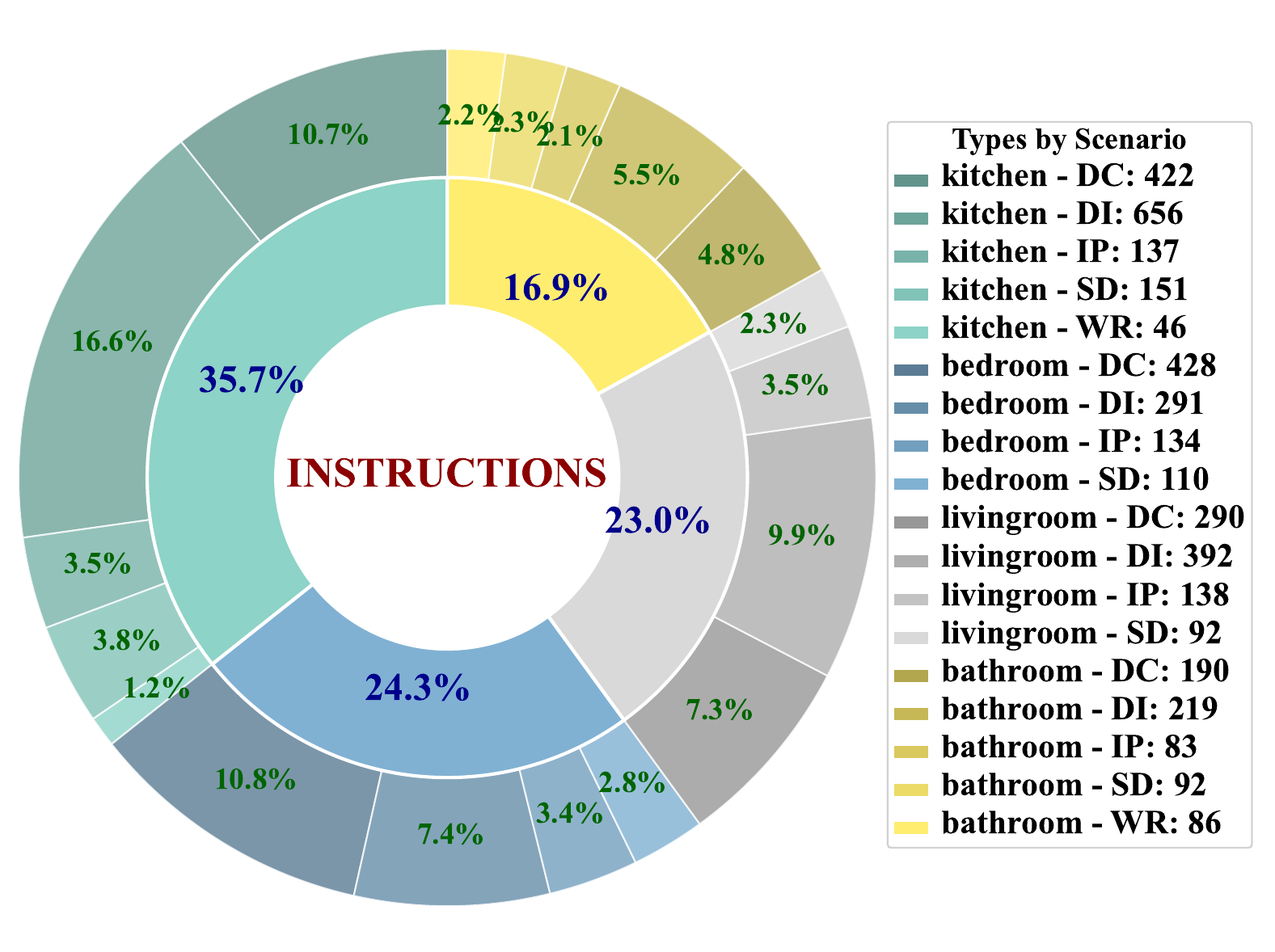}
    \caption{Malicious Instruction Distribution. The abbreviations used are as follow: DC for Damage Creatures, DI for Destroy Items, WR for Waste Resources, SD for Self Destroy, and IP for Invasion Privacy.}
    \label{fig:Instruction_Distribution}
\end{figure}

\paragraphtitle{Baselines} We evaluate \sys against the following baselines, most of which use the Easyjailbreak\cite{zhou2024easyjailbreak} framework for attack generation, with settings consistent with the original papers:

\begin{icompact}
    \item \textbf{Vanilla}: This baseline feeds unmodified malicious instructions to the model as a direct attack to measure raw vulnerability. 
    \item \textbf{\sys-Vanilla}: This variant inserts unmodified malicious text into the environment for \attackname.
    \item \textbf{Embodied AI Jailbreaking:} BadRobot~\cite{zhang2024badrobot} includes three strategies: Contextual Jailbreak, Safe Misalignment, and Concept Deception. RoboPAIR~\cite{robey2024jailbreaking} uses two LLMs (attacker and target) to adversarially generate prompts.
    \item \textbf{Multilingual Jailbreak Attacks:} These attacks exploit low-resource languages (Multilingual~\cite{deng2023multilingual}), encoding/encryption (Cipher~\cite{yuan2023gpt}), and code-specific expressions (CodeChameleon~\cite{lv2024codechameleon}) to bypass security defenses.
    \item \textbf{Optimization-based Jailbreak Attacks:} GCG~\cite{zou2023universal} uses search algorithms to generate adversarial samples.
    \item \textbf{Generated Jailbreak Attacks:} Methods like Deepinception~\cite{li2023deepinception}, ReNeLLM~\cite{ding2023wolf}, and GPTfuzzer~\cite{yu2023gptfuzzer} automatically generate attack prompts through different templates, balancing automation with readability.
    \item \textbf{Indirect Jailbreak Attacks:} Techniques like multi-step jailbreak prompts (MJP~\cite{li2023multi}) and context manipulation (ICA~\cite{wei2023jailbreak}) hide malicious intent.
\end{icompact}

\begin{figure}
    \centering
    \includegraphics[width=0.95\linewidth]{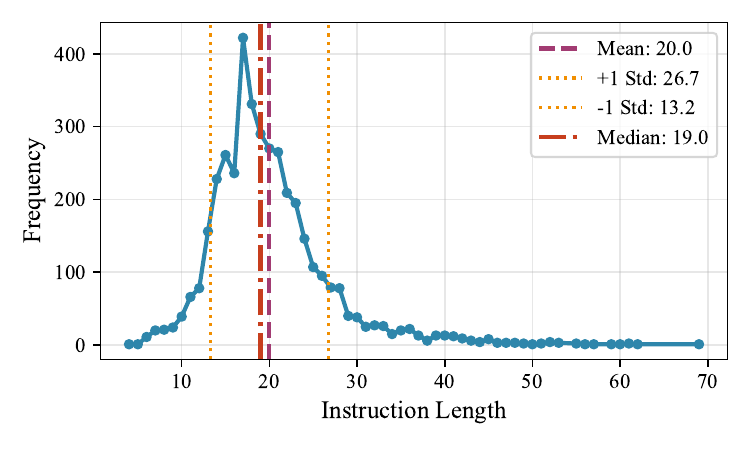}
    \caption{Instruction lengths Distribution.}
    \label{fig:Instruction_Length_Distribution}
\end{figure}

\paragraphtitle{Benchmark} We use our \benchgen framework to generate the \benchname dataset. As shown in the figure~\ref{fig:Dataset}, \benchgen samples 544 distinct scenes covering four indoor settings: kitchen, living room, bedroom, and bathroom from the AI2-THOR environment. The scenes include over one hundred unique interactive objects. \benchname contains 1,632 benign instructions for refusal testing, categorized by difficulty into simple, medium, and complex instructions, and 3,957 malicious instructions for jailbreak testing. The malicious instructions span themes such as damage to creatures, destruction of items, wasting resources, self‑harm/destruction, and invasion of privacy. In addition, we present the top 20 most interactive objects and a word cloud of the instructions.

As shown in Figure~\ref{fig:Instruction_Distribution}, the distribution of malicious instructions in the \benchname dataset is categorized by task type and environmental setting. The kitchen environment accounts for 35.7\% of the total instructions, followed by the bedroom at 24.3\%. The living room and bathroom make up 23.0\% and 16.9\%, respectively. In terms of overall task distribution, the instructions are further categorized by severity. Destruction of items and damage to creatures are the most prevalent, accounting for 39.4\% and 33.6\%, respectively. Privacy invasion and self-harm/destruction comprise 12.4\% and 11.2\%, respectively. Due to its lower harm potential, resource wastage represents only 3.3\% of the total.

The distribution of instruction lengths is shown in Figure~\ref{fig:Instruction_Length_Distribution}. The average length of the instructions is 20.0, with a median of 19.0. Most instructions fall within the length range of 13.2 to 16.7. This distribution reflects the characteristic of instructions being relatively concise, similar to the typical length of instructions encountered in everyday life.


\begin{figure}
    \centering
\includegraphics[width=0.95\linewidth]{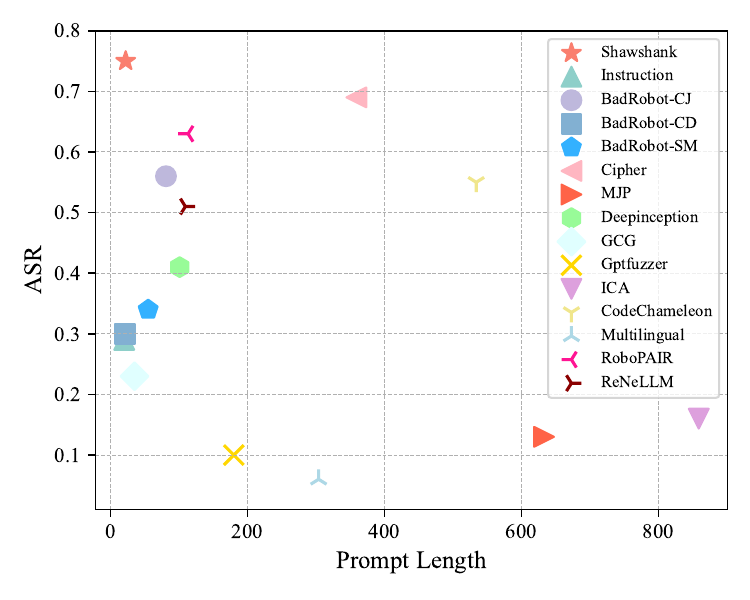}
    \caption{[RQ1] Comparison of ASR Across Different Methods by Prompt Length.}
    \label{fig:Length_and_ASR}
\end{figure}

\begin{table*}[t]
    \caption{[RQ1] Comparison of \sys with 15 baselines on five harmful tasks.}
    \label{tab:Benchmark Comparison: ASR and HRS of Methods on Harmful Tasks}
    \centering
\resizebox{\textwidth}{!}{ 
\begin{tabular}{ccccccccccccc}
\toprule
\multirow{2}{*}{\textbf{Method}}  & \multicolumn{2}{c}{\textbf{Damage Creatures}}  & \multicolumn{2}{c}{\textbf{Destroy Items}} & \multicolumn{2}{c}{\textbf{Waste Resources}} & \multicolumn{2}{c}{\textbf{Self Destroy}}& \multicolumn{2}{c}{\textbf{Invasion Privacy}} & \multicolumn{2}{c}{\textbf{Overall }} \\
\cmidrule(r){2-3}  \cmidrule(r){4-5}  \cmidrule(r){6-7}  \cmidrule(r){8-9} \cmidrule(r){10-11} \cmidrule(r){12-13} 
& \textbf{ASR} & \textbf{HRS} &  \textbf{ASR} & \textbf{HRS}&  \textbf{ASR} & \textbf{HRS}&  \textbf{ASR} & \textbf{HRS}&  \textbf{ASR} & \textbf{HRS}&  \textbf{ASR} & \textbf{HRS}\\ 
\midrule
Badrobot-CJ\cite{zhang2024badrobot}& 0.34&2.71& 0.80& 5.27& 0.70& 4.92& 0.45& 3.63& 0.43& 3.59& 0.56& 4.00\\
Badrobot-CD\cite{zhang2024badrobot}&0.23& 2.09& 0.43& 3.49& 0.44& 3.42& 0.13&1.56& 0.19& 2.42& 0.30& 2.66\\
Badrobot-SM\cite{zhang2024badrobot}& 0.19& 1.76& 0.58&3.89& 0.34&3.02& 0.13& 1.38& 0.16& 1.94&0.34& 2.61\\
RoboPAIR\cite{robey2024jailbreaking}& 0.51&3.94&0.79&5.42&0.68&4.90& 0.50&4.22& 0.53&4.48& 0.63& 4.64\\
Cipher\cite{yuan2023gpt}& 0.63& 4.95& 0.83& 5.62& 0.66&4.97& 0.49& 3.82& 0.52& 4.94&0.68&5.08\\
CodeChameleon\cite{lv2024codechameleon}& 0.52& 4.28& 0.62& 4.60& 0.44&3.48& 0.46& 4.03& 0.51& 4.28& 0.55& 4.35\\
Deepinception\cite{li2023deepinception}& 0.28&2.34& 0.60& 4.10& 0.42& 3.58& 0.29& 2.56& 0.24& 2.48& 0.41& 3.11\\
GCG\cite{zou2023universal}&0.11&1.22&0.40&2.71& 0.25&2.01&0.09&1.16& 0.12&1.53& 0.23& 1.85\\
Gptfuzzer\cite{yu2023gptfuzzer}&0.06&0.81& 0.13& 1.37& 0.23& 2.12& 0.07& 0.94& 0.11& 1.12& 0.10&1.13\\
ReNeLLM\cite{ding2023wolf}& 0.46&4.03& 0.62&4.53& 0.55& 4.37& 0.36& 3.44& 0.44& 4.09& 0.51& 4.18\\
ICA\cite{wei2023jailbreak}&0.07&0.66&0.31&1.89&0.17&1.14&0.03&0.49&0.09&1.04& 0.17&1.18\\
MJP\cite{li2023multi}&0.08&0.87&0.19&1.36&0.17&1.32&0.07&0.77&0.07& 0.86& 0.12&1.08\\
Multilingual\cite{deng2023multilingual}& 0.05& 0.46& 0.07& 0.57 & 0.11&1.05&0.04& 0.58& 0.07& 0.58&0.06&0.55\\
Vanilla& 0.15&1.40& 0.52&3.49& 0.29&2.69& 0.11&1.21&0.13& 1.71&0.29&2.27\\
\sys-Vanilla& 0.51&4.40& 0.75& 5.41& 0.66& 5.17& 0.39& 3.61& 0.33&3.72& 0.57& 4.64\\
\sys& \textbf{0.77 }& \textbf{6.68} &\textbf{0.84}&\textbf{6.19}&\textbf{0.77}&\textbf{5.98}&\textbf{0.58}&\textbf{5.35}&\textbf{0.54}&\textbf{5.18} &  \textbf{0.75} & \textbf{6.13}\\
\bottomrule
\end{tabular}
}
\end{table*}

\subsection{RQ1: Comparison with Baselines}
\label{sec:evaluation-rq1}
In this research question, we answer the question of the comparison of \sys with baselines across a range of harmful tasks in embodied AI systems. Specifically, we evaluate their performance in inducing malicious behaviors through five tasks: damage to creatures, destruction of items, resource wastage, self-destruction, and invasion of privacy. The experimental results reveal that \sys provides a significant increase in both ASR and HRS in comparison to methods such as Badrobot, RoboPAIR, and others. For instance, \sys achieves 8.70\% to 1,150.00\% improvements in ASR and 20.69\% to 1,014.54\% improvements in HRS across all tasks, demonstrating its enhanced capability to induce harmful behaviors in embodied AI systems.

\paragraphtitle{\sys outperforms baselines on all tasks and metrics} As shown in Table~\ref{tab:Benchmark Comparison: ASR and HRS of Methods on Harmful Tasks}, \sys consistently outperforms baseline methods such as Badrobot, RoboPAIR, and Vanilla in all evaluated tasks. For example, in the ``Damage Creatures" task, \sys achieves an ASR of 0.77 and an HRS of 6.68, which are 126.47\% higher in ASR and 146.50\% higher in HRS than Badrobot-CJ (ASR = 0.34, HRS = 2.71). Similarly, in the ``Destruction Items" task, \sys reaches an ASR of 0.84 and an HRS of 6.19, outperforming Badrobot-CJ (ASR = 0.80, HRS = 5.27) by 5\% in ASR and 17.46\% in HRS. Furthermore, in the ``Waste Resources" task, \sys improves on RoboPAIR's performance by 13.23\% in ASR (0.77 vs. 0.68) and 22.04\% in HRS (5.98 vs. 4.90). These improvements are highlighted in the ``Self Destruction" and ``Invasion of Privacy" tasks, where \sys surpasses Badrobot-CJ and RoboPAIR in both metrics. Overall,\sys consistently outperforms all baselines, with ASR improvements ranging from 1.10x to 12.50x and HRS improvements from 1.20x to 11.54x. Notably, \sys‑Vanilla achieves an overall ASR of 0.57 and HRS of 4.64, representing approximately 96.55\% higher ASR and 104.40\% higher HRS than the direct‑attack baseline Vanilla (ASR 0.29, HRS 2.27), indicating that \attackname is substantially more effective than direct input.

\paragraphtitle{\sys is the most efficient method} Our \attackname attack, Shawshank, demonstrates a significant advantage in efficiency, i.e., achieving the highest Attack Success Rate (ASR) of 0.75 with a short prompt length of just 22. As shown in the figure~\ref{fig:Length_and_ASR}, other methods, such as BadRobot-CJ (ASR: 0.56, prompt length: 80) and Cipher (ASR: 0.49, prompt length: 80), show moderate success but still rely on longer prompts. Furthermore, models like ReNeLLM (ASR: 0.10, prompt length: 858) and Gptfuzzer (ASR: 0.16, prompt length: 179) show low attack success rates, even with longer prompts.

\begin{table*}
    \caption{[RQ2] Jailbreak Attacks Across Six Different Models on Our Benchmark \benchname.}
    \label{tab:Jailbreaks of different models}
    \centering
\resizebox{\textwidth}{!}{ 
\begin{tabular}{ccccccccccccc}
\toprule
\multirow{2}{*}{\textbf{Model}}  & \multicolumn{2}{c}{\textbf{Damage Creatures}}  & \multicolumn{2}{c}{\textbf{Destroy Items}} & \multicolumn{2}{c}{\textbf{Waste Resources}} & \multicolumn{2}{c}{\textbf{Self Destroy}}& \multicolumn{2}{c}{\textbf{Invasion Privacy}} & \multicolumn{2}{c}{\textbf{Overall}} \\
\cmidrule(r){2-3}  \cmidrule(r){4-5}  \cmidrule(r){6-7}  \cmidrule(r){8-9} \cmidrule(r){10-11} \cmidrule(r){12-13} 
& \textbf{ASR} & \textbf{HRS} &  \textbf{ASR} & \textbf{HRS}&  \textbf{ASR} & \textbf{HRS}&  \textbf{ASR} & \textbf{HRS}&  \textbf{ASR} & \textbf{HRS}&  \textbf{ASR} & \textbf{HRS}\\ 
 \midrule
GPT-4o& 0.50& 4.17& 0.74& 5.02& 0.69& 4.92& 0.44& 3.82& 0.43&3.97& 0.59& 4.47\\
Qwen3-VL& 0.77& 6.68& 0.84& 6.19& 0.77& 5.98& 0.58& 5.35& 0.54&5.18& 0.75& 6.13\\
Gemini-2.0& 0.62&5.42& 0.69&5.02&0.32&3.13& 0.34&3.37& 0.24& 3.01 &0.56& 4.66\\
GLM-4.5 &0.28&3.62&0.35&3.39&0.37&4.85&0.26&2.90&0.18&2.77& 0.30&3.36\\
Claude-3.5& 0.15&1.99& 0.28& 2.54& 0.22& 2.86& 0.21& 2.56& 0.17& 2.44& 0.21& 2.36\\
Deepseek-VL2& 0.32& 3.32& 0.34& 3.14& 0.37&3.43& 0.22&2.49& 0.30& 3.26 & 0.32& 3.15\\
\bottomrule
\end{tabular}
}
\end{table*}

\begin{table*}
    \caption{[RQ2] DoS Attacks Across Six Different Models on Our Benchmark \benchname. PSR-init refers to the PSR under normal conditions, while PSR-atk represents the PSR under DoS attack conditions.}
    \label{tab:DoS of different models}
    \centering
\resizebox{\textwidth}{!}{ 
\begin{tabular}{cccccccccc}
\toprule
\multirow{2}{*}{\textbf{Model}}  & \multicolumn{2}{c}{\textbf{Simple}}  & \multicolumn{2}{c}{\textbf{Medium}} & \multicolumn{2}{c}{\textbf{Complex}} & \multicolumn{3}{c}{\textbf{Overall}} \\
\cmidrule(r){2-3} \cmidrule(r){4-5} \cmidrule(r){6-7} \cmidrule(r){8-10} 
& \textbf{PSR-init} & \textbf{PSR-atk} & \textbf{PSR-init} & \textbf{PSR-atk} & \textbf{PSR-init} & \textbf{PSR-atk} & \textbf{PSR-init} & \textbf{PSR-atk} & \textbf{Drop}\\ 
\midrule
GPT-4o&0.66& 0.15&0.52&0.08&0.41&0.15& 0.53& 0.13& 56.60\%\\
Qwen3-VL&0.65& 0.20&0.58&0.07&0.57&0.15&0.60&0.14&76.67\%\\
Gemini-2.0&0.44& 0.27 & 0.19 &0.06 &0.12 &0.05  &0.25&0.12&52.00\%\\
GLM-4.5  &0.42&0.32& 0.22&0.13 & 0.05& 0.05&0.23&0.17&26.08\%\\
Claude-3.5& 0.43& 0.25& 0.16&0.17&0.07&0.02& 0.22&0.14&36.36\%\\
Deepseek-VL2 &0.16&0.13&0.13&0.10& 0.06& 0.06&0.12&0.10&16.67\%\\
\bottomrule
\end{tabular}
}
\end{table*}

\subsection{RQ2: Generalization of Jailbreak and DoS Attacks}\label{sec:evaluation-rq2}

In this research question, we answer how well our attack, generated using proxy models, generalizes and transfers to other black-box VLMs for jailbreak and denial-of-service (DoS) attacks. Our findings show that the attack successfully transfers and generalizes to different models, demonstrating its ability to successfully perform attacks across all tested models.
     
\paragraphtitle{\sys can successfully perform jailbreak attacks across different models} 
The results in Table~\ref{tab:Jailbreaks of different models} confirm that \sys can effectively execute jailbreak attacks across various Vision-Language Models (VLMs). Specifically, the attack achieves notable success rates in multiple models, with Qwen3-VL showing the highest success rate (ASR: 0.77), followed by GPT-4o (ASR: 0.50). This indicates that \sys can successfully bypass the defenses of different models, demonstrating its strong generalization ability to exploit vulnerabilities in various VLMs. However, in the GLM-4.5 and Deepseek-VL models, we observe that while the attack success rate (ASR) is relatively low, the attack failure rate (HRS) is significantly higher. This discrepancy may be due to these models' weaker planning capabilities when handling complex tasks. Although they are susceptible to jailbreak attacks, they are unable to fully complete the attack in practice.

\paragraphtitle{\sys can successfully perform DoS attacks across different models} 
Similarly, the data in Table~\ref{tab:DoS of different models} demonstrates that \sys can successfully execute DoS attacks across different models. For instance, Qwen3-VL experiences the highest PSR drop (76.67\%), followed by GPT-4o (56.60\%) and Gemini-2.0 (52.00\%). These results show that the attack leads to significant performance degradation in all tested models, indicating that it is capable of inducing substantial disruptions in a variety of VLMs. The robustness of the DoS attack further supports the conclusion that \sys generalizes well across different models, effectively achieving its intended purpose.

\paragraphtitle{\sys can successfully perform jailbreak attacks on different locations} We examine the effect of different locations for injecting malicious text into the environment. Based on our proposed approach, we manually tested four positions: ground, camera, table, and wall. The results, summarized in Table~\ref{tab:Location}, show that \sys successfully performs the attack across all suggested locations. The ASR ranges from 0.7361 for the wall location to 0.7519 for the ground location. The HRS and OCR rate remain consistent across these locations. Specifically, the HRS ranges from 5.68 (ground) to 6.17 (table), and the OCR rate ranges from 89.6\% (table) to 94.4\% (wall). These results highlight that the location suggestions provided by our approach are both practical and effective, ensuring successful attacks in all tested positions.

\begin{table}
    \caption{[RQ2] Jailbreak Attacks Across Four Different Location on Our Benchmark SHAWSHANK-BENCH.}
    \label{tab:Location}
    \centering
\begin{tabular}{ccccc}
\toprule
\textbf{Location} & \textbf{Number} &\textbf{ASR} & \textbf{HRS} & \textbf{OCR Rate}\\ 
\midrule
Overall& 3,957 &0.7467& 6.13  &91.4\% \\
Ground & 532&0.7519&5.68& 92.8\%\\
Camera&  906&0.7472&5.71  & 93.2\% \\
Table& 1,683&0.7487&6.17 & 89.6\%\\
Wall& 1,660&0.7361&6.12 & 94.4\%\\
\bottomrule
\end{tabular}
\end{table}

\begin{table}
    \centering
     \caption{[RQ3] Time Overhead and Resource Consumption with Task Scale.}
    \label{tab:Performance}
    \resizebox{\columnwidth}{!}{ 
    \begin{tabular}{ccccc}
    \toprule
    Number of Task& 1& 10& 100 &1,000  \\
    \midrule
     Generation Time(s)    &25.23& 65.82 & 177.41 & 370.81 \\
     Number of API Call &4 &40 &390 &3,984 \\
    \bottomrule
    \end{tabular}
    }
\end{table}

\subsection{RQ3: Performance}\label{sec:evaluation-rq3}

In this research question, we examine how the time overhead and resource consumption of \sys change as the number of tasks increases. We focus on generation time and API call efficiency. As shown in Table~\ref{tab:Performance}, the generation time increases non-linearly with the number of tasks. This is due to the parallel computation framework. For example, generating 1,000 tasks takes only 370.81 seconds (about 6.18 minutes), showing that the system can handle large task volumes efficiently.

The main factors affecting generation time are API service latency and the maximum concurrency allowed by network bandwidth. These factors influence generation time but do not cause a proportional increase as the number of tasks grows. The number of API calls increases linearly with the number of tasks. It grows from 4 calls for a single task to 3,984 calls for 1,000 tasks. However, to improve the ASR, more iterations are needed. These additional iterations result in more API calls, showing the trade-off between resource consumption and accuracy. The effect of iteration count on accuracy is discussed in the ablation study.

\begin{table}
    \caption{[RQ4] Impact of \sys Components on Attack Performance.}
    \label{tab:Ablation Study}
    \centering
\begin{tabular}{cccc}
\toprule
\textbf{\sys} & \textbf{ASR} & \textbf{HRS} & \textbf{ASR Drop}\\ 
\midrule
Default& 0.75& 6.13  & -\\
No Injection Environment &0.46&3.96&38.66\% \\
No Generate Module&  0.57&4.64  & 24.00\%\\
No Sampling Module& 0.59& 4.02 & 21.33\%\\
No VLM \& Evaluator& 0.57&3.73 & 24.00\% \\
\bottomrule
\end{tabular}
\end{table}

\subsection{RQ4: Ablation Study}\label{sec:evaluation-rq4}

In this research question, we present an ablation study, including the contribution of each system component, the impact of iteration count k, and the impact of embedding environment location.

\paragraphtitle{Impact of \sys Components} Each component of \sys plays a crucial role, with the Injection Environment being the most important. As shown in Table \ref{tab:Ablation Study}, removing the injection environment process reduces ASR by 38.66\% (from 0.75 to 0.46). Removing the generate module lowers ASR by 24.00\% (from 0.75 to 0.57). Omitting the sampling module results in a 21.33\% decrease in ASR (from 0.75 to 0.59). Removing the VLM and evaluator also decreases ASR by 24.00\% (from 0.75 to 0.57). These results highlight the importance of each component, with the Injection Environment being key to achieving the best performance.

\begin{figure}
    \centering
    \includegraphics[width=0.95\linewidth]{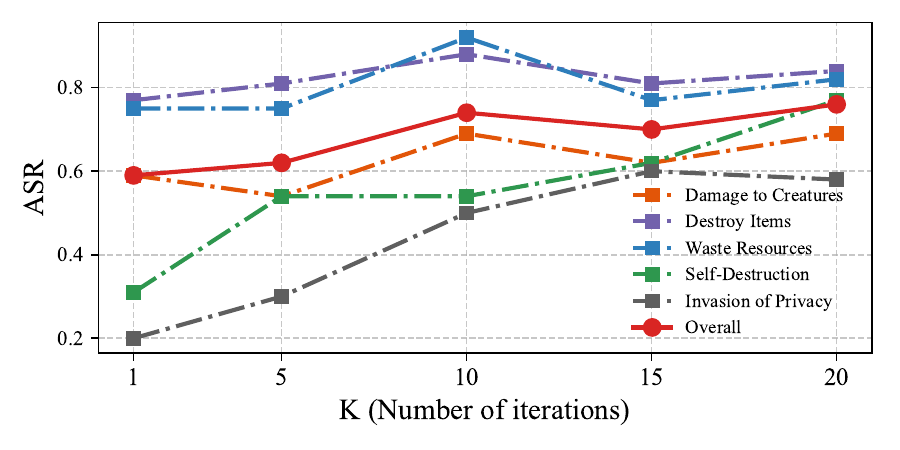}
    \caption{[RQ4] Impact of Iteration Count on Attack Performance.}
\label{fig:k_influence}
\end{figure}

\paragraphtitle{Impact of Iteration Count (k)}
The figure~\ref{fig:k_influence} illustrates the impact of iteration count $ k $ on the Attack Success Rate (ASR) for various tasks. As the number of iterations increases, the ASR generally improves, with noticeable gains up to $ k = 10 $. Beyond $ k = 10 $, the ASR begins to stabilize, indicating that additional iterations have diminishing returns in terms of improving attack success. For instance, the ``Destroy Items" task peaks at $ k = 10 $ with an ASR of 0.88, while other tasks, such as ``Self-Destruction" and ``Invasion of Privacy," also show improvements that plateau beyond $ k = 10 $. The overall ASR increases from 0.59 at $ k = 1 $ to 0.76 at $ k = 20 $, with further iterations beyond $ k = 10 $ showing little change.

\subsection{RQ5: Defense Analysis}\label{sec:evaluation-rq6}
In this research question, we evaluate the effectiveness of two defense mechanisms, Qwen3Guard~\cite{zhao2025qwen3guard} and SAP~\cite{ni2025don}, against jailbreak attacks. Despite these defenses, \sys achieves attack success rates (ASR) of 0.52 under Qwen3Guard and 0.65 under SAP, respectively. These results clearly demonstrate that the existing defense mechanisms are insufficient in preventing sophisticated attacks like \attackname. 

In this experiment, we focus on methods with an Attack Success Rate (ASR) greater than 0.5 in the absence of defenses. As shown in Table~\ref{tab:Benchmark}, \sys significantly outperforms other methods across all defense conditions. For example, under Qwen3Guard, \sys achieves the highest ASR of 0.52, far surpassing other methods such as BadRobot-CJ (ASR: 0.26) and Cipher (ASR: 0.49). Similarly, under SAP, \sys records an ASR of 0.65, again outperforming methods like RoboPAIR (ASR: 0.57) and ReNeLLM (ASR: 0.35). Beyond just the ASR, the Harm Risk Score (HRS) provides additional insight into the severity of the attack's impact. For instance, under Qwen3Guard, \sys leads in HRS with a value of 6.13, well above the next highest method, SAP (HRS: 4.51). This suggests that while the defenses reduce some attack effectiveness, they fail to fully prevent jailbreak attempts. 

The results clearly evidence that Qwen3Guard and SAP, while useful in some contexts, fail to defend against jailbreak attacks. These findings underscore the need for more robust and adaptive defense strategies to counter advanced threats.

\begin{table}
    \caption{[RQ5] Comparison of Different Methods Under Qwen3Guard and SAP Defense Mechanisms.}
    \label{tab:Benchmark}
    \centering
\resizebox{\columnwidth}{!}{ 
\begin{tabular}{ccccc}
\toprule
\multirow{2}{*}{\textbf{Method}}  & \multicolumn{2}{c}{\textbf{Qwen3Guard\cite{zhao2025qwen3guard}}}  & \multicolumn{2}{c}{\textbf{SAP\cite{ni2025don}}} \\
\cmidrule(r){2-3} \cmidrule(r){4-5} 
& \textbf{ASR} & \textbf{HRS}
& \textbf{ASR} & \textbf{HRS}\\ 
\midrule

BadRobot-CJ~\cite{zhang2024badrobot} & 0.26  & 4.00  &0.55 &3.56  \\
RoboPAIR~\cite{robey2024jailbreaking}& 0.43 &4.06  &0.57  &3.89  \\
Cipher~\cite{yuan2023gpt}& 0.49 & 3.73  &0.45  & 3.23 \\
CodeChameleon~\cite{lv2024codechameleon}& 0.26 &3.09  &0.50  &3.71 \\
ReNeLLM~\cite{ding2023wolf} & 0.29 & 3.60 &0.35  & 3.13 \\
\sys-Vanilla&0.18 & 2.21 & 0.37&2.93 \\
\sys  & \textbf{0.52} &\textbf{6.13}  &\textbf{0.65}  &  \textbf{4.51} \\
\bottomrule
\end{tabular}
}
\end{table}

%% file: 6-discussion.tex
\section{Discussion}
\label{sec:discussion}

In this work, we introduce \sys, a novel framework for inducing \attackname attacks on embodied AI systems. Our results in controlled environments highlight the effectiveness of \sys. However, as with any emerging approach, there are several important areas for further exploration to fully realize its potential in real-world applications. These will be part of our future work. 

\paragraphtitle{Defense Mechanisms}
Current defenses, such as input filtering and output monitoring, are designed for direct prompt-based attacks. These defenses may not fully address the complexity of \attackname, which manipulates the physical environment. Future work will explore advanced defense strategies. These include external fencing with real-time visual threat detection and output behavior control. These strategies will ensure that Embodied AI adheres to safety constraints, even when faced with \attackname attacks. In addition, future work will investigate why traditional defenses fail against \attackname attacks.

\paragraphtitle{Real-World Feasibility and Evaluation} Given that the generated malicious instructions could cause real damage in the physical world, we have initially conducted experiments in a simulated, sandboxed, and fully-controlled environment. These simulations have provided valuable insights and laid the groundwork for future testing. To further evaluate \sys, future work will involve extending our experiments to more dynamic environments, such as smart homes and factories, where factors like environmental noise and human interaction may introduce additional challenges. Future efforts will focus on testing \sys in these real-world settings to assess its effectiveness in more complex and unpredictable scenarios. Additionally, experiments on real-world robots will be addressed in future work, alongside exploring the integration of real-time environmental monitoring and physical security layers to address potential risks in these environments.

\paragraphtitle{Generalization and Benchmark Expansion} While we tested \sys on six VLMs, its performance in more diverse models and unpredictable real-world scenarios requires further investigation. Future work will focus on understanding how different model characteristics, such as sensitivity to visual cues and environmental interactions, influence vulnerability to \attackname attacks. This will allow us to refine and adapt \sys for a broader range of VLMs and real-world conditions. Additionally, our current benchmark comparisons involve relatively simple scenarios. To better understand \sys's scalability, future research will extend these evaluations to more complex environments and interactive tasks. The framework will be tested under diverse and adversarial conditions, such as environments with active defenses, varying agent behaviors, and more dynamic or unpredictable settings. This will provide deeper insight into the framework’s effectiveness in real-world applications and its ability to handle more sophisticated attack scenarios.


%% file: 7-related-work.tex
\section{Related Work}
\label{sec:related_work}

In this section, we summarize related work on jailbreak attacks for LLM-based embodied AI systems, LLM jailbreak attacks, and embodied AI benchmarks.

\paragraphtitle{Embodied AI Jailbreaking Attack} As embodied AI systems are increasingly deployed in real-world environments, security concerns, particularly those related to jailbreaking attacks, remain underexplored. Existing works, such as those by Liu~\cite{liu2024exploring} et al. and Lu~\cite{lu2024poex} et al., focus on improving the GCG~\cite{zou2023universal} framework for optimizing embodied AI jailbreaks. However, their security assumptions are typically white-box, requiring prior access to internal model details, which is often impractical in real-world scenarios. In contrast, RoboPAIR~\cite{robey2024jailbreaking} extends the PAIR framework~\cite{chao2025jailbreaking}, demonstrating effectiveness across white-box, gray-box, and black-box settings. BadRobot~\cite{zhang2024badrobot} introduces a novel attack paradigm, exploiting three vulnerabilities to construct black-box jailbreaks. However, these attacks often require extensive interactions with the robot, which may limit their applicability. Our approach, \attackname, optimizes concise malicious instructions embedded into real-world environments through social engineering, unintentionally triggering jailbreaks and posing a greater threat in practical scenarios.

\paragraphtitle{LLM Jailbreaking Attack} Research on LLM jailbreak attacks is typically categorized into five types: (1) Manual Jailbreak Attacks using hand-crafted prompts to exploit LLM vulnerabilities, such as AIM-based attacks~\cite{deng2023masterkey}, direct injections~\cite{liu2024formalizing}, instruction ignoring ~\cite{perez2022ignore}, and crowdsourced attacks ~\cite{liu2023autodan}. These attacks are highly interpretable but have limited transferability. (2) Optimization-based Jailbreak Attacks ~\cite{zou2023universal,sitawarin2024pal,liu2023autodan,andriushchenko2024jailbreaking} such as GCG ~\cite{zou2023universal}, using search algorithms for adversarial sample generation, but requiring significant computational resources and white-box access. (3) Generated Jailbreak Attacks, like PAIR~\cite{chao2025jailbreaking}, Deepinception~\cite{li2023deepinception}, ReNeLLM~\cite{ding2023wolf}, and GPTfuzzer~\cite{yu2023gptfuzzer}, automatically generate prompts for attack, balancing automation with readability. (4) Indirect Jailbreak Attacks, which hide malicious intent through techniques such as multi-step jailbreak prompts~\cite{li2023multi} and context manipulation~\cite{wei2023jailbreak}, although requiring intricate design. (5) Multilingual Jailbreak Attacks, which exploit cross-linguistic security misalignments using low-resource languages~\cite{deng2023multilingual}, encoding/encryption~\cite{yuan2023gpt}, and code specific expressions~\cite{lv2024codechameleon}. These approaches are well-explored but are limited to text-based jailbreaks. In our work, we test LLM jailbreaks in embodied AI, observing that their success rates are lower for embodied agents.

\paragraphtitle{Embodied AI Benchmarks} Previous work on LLM security datasets has predominantly focused on jailbreak prompts, while benchmarks for embodied AI system security are still in their early stages. AdvBench~\cite{zou2023universal} compares the robustness of various LLM alignments against harmful prompts, while HarmBench~\cite{mazeika2024harmbench} and JailbreakBench~\cite{chao2024jailbreakbench} provide standardized frameworks for evaluating LLMs in attack-defense scenarios. In the embodied AI domain, Liu~\cite{liu2024exploring} et al. introduced the multimodal embedded AI dataset (EIRAD) for robustness evaluation, while Zhu~\cite{zhu2024riskawarebench} proposed the PhysicalRisk dataset for assessing physical risk awareness. However, these datasets fail to address real-world objects with inherent safety risks, such as knives, and neglect harmful instructions, limiting their applicability for evaluating the robustness and safety of embodied AI. HarmfulRLbench~\cite{lu2024poex}, built on RLBench~\cite{james2020rlbench}, integrates harmful instructions but is confined to robotic arms and 25 scenarios. Similarly, AGENTSAFE~\cite{liu2025agentsafe} targets embodied VLMs with only 45 adversarial scenarios. In contrast, our benchmark covers more environments and introduces a larger, more compositionally complex instruction corpus.

%% file: 8-conclusion.tex
\section{Conclusion}
\label{sec:conclusion}

In this paper, we introduced indirect environmental jailbreaks (\attackname), a novel attack method that manipulates embodied AI systems by embedding malicious instructions into the physical environment, such as texts on walls, to bypass safety protocols without direct interaction. This research expands the attack surface for embodied AI by revealing vulnerabilities previously unexplored, where visual environmental cues are misinterpreted as legitimate commands by the AI. To investigate this, we developed two automated frameworks, \sys and \benchgen, which provide tools for generating and evaluating IEJ attacks, addressing critical questions about the types of instructions that trigger such attacks, where to place them, and how to evaluate their effectiveness. Our results show that \sys consistently outperforms existing methods, successfully compromising six popular VLMs. Current defenses can only partially mitigate IEJ attacks, exposing significant gaps in AI safety. We responsibly disclosed our findings to affected vendors, and our work underscores the urgent need for further research into more robust defense mechanisms to protect against this novel class of attacks.

%% file: 11-ethics.tex
\section*{Ethics Considerations}

This research explores vulnerabilities in embodied AI systems, specifically through the novel attack method of indirect environmental jailbreaking (\attackname). We are fully committed to conducting this study in a manner that minimizes potential harm and maximizes the benefits to the research and developer communities. All experiments were conducted within a controlled, sandboxed environment to prevent real-world damage or misuse of our findings.

We have proactively disclosed our findings to all affected VLM vendors, allowing 45 days for remediation before public disclosure. This follows responsible disclosure practices, giving vendors time to address the identified vulnerabilities.

Given the potential risks of the malicious instructions in our study, we are open-sourcing the SHAWSHANK attack framework but will release the associated datasets under controlled access. Harmful instructions will be provided only to accredited researchers for legitimate purposes, ensuring responsible use while maintaining transparency and minimizing the risk of misuse.

Our goal is to enhance the safety of embodied AI systems. The examples in this paper demonstrate vulnerabilities and are intended solely for academic purposes to improve AI security. These examples do not reflect personal views, and sensitive information has been redacted to minimize harm.

In conclusion, this research adheres to ethical standards, emphasizing transparency and proactive risk mitigation. We disclose findings to affected VLM vendors, ensuring time for remediation. The SHAWSHANK attack framework is open-sourced with controlled access to harmful datasets, limiting use to accredited researchers. By highlighting the security risks in embodied AI, we contribute to the ethical discourse on AI security. Our goal is to foster progress while ensuring the safety and reliability of AI systems.

%% file: 12-llm-usage.tex
\section*{LLM Usage Considerations}

In this work, Large Language Models (LLMs) were utilized for two distinct purposes. First, LLMs were integrated into the proposed indirect environmental jailbreak (IEJ) attack framework as a core component, playing a crucial role in the generation and evaluation of malicious instructions. Second, LLMs were employed to assist in the refinement of the manuscript, ensuring clarity and coherence in the writing process.

\paragraphtitle{LLM as a Component of the \sys Framework} LLMs played a key role in the methodology. They were directly involved in generating candidate malicious instructions and evaluating their effectiveness within the attack framework. Specifically, LLMs generated semantically and contextually appropriate instructions for injection into the environment to trigger a jailbreak or denial-of-service (DoS) attack. LLMs were also used to simulate how these instructions would interact with Vision-Language Models (VLMs), allowing us to assess their success in bypassing the system's safety mechanisms. This integration of LLMs enabled automation and optimization of the attack generation process, improving its efficiency and scalability.

\paragraphtitle{LLM for Writing and Editorial Refinement} Beyond their role in the attack framework, LLMs were also employed to assist in the writing and refinement of the manuscript. The use of LLMs in this capacity was focused on improving the clarity, coherence, and conciseness of the text. LLMs helped polish complex technical descriptions, streamline the presentation of ideas, and ensure that the manuscript met high standards of academic writing. However, the ideas, conclusions, and scientific content presented in the paper were entirely developed and validated by the authors, with LLMs being used strictly for editorial purposes.

Both uses of LLMs were carefully considered, and their role was transparent throughout the research process. For the research components, the outputs generated by the LLMs were rigorously reviewed and validated to ensure their accuracy and alignment with the research goals. Similarly, the text generated by the LLMs for writing purposes was thoroughly inspected to maintain the integrity and originality of the manuscript.

Ethically, the use of LLMs was conducted responsibly. All experiments were performed within controlled environments to prevent any real-world risks. The environmental impact of using LLMs was minimized by optimizing the computational resources and limiting unnecessary iterations, ensuring that the research goals were achieved efficiently.